\documentclass[aps,superscriptaddress,twocolumn,floats,prd]{revtex4} 
\usepackage[svgnames]{xcolor}
\usepackage[colorlinks=true, linkcolor=blue!60!black, citecolor=blue!60!black, linktoc=all]{hyperref}
\usepackage{graphicx} %
\usepackage{epsfig,amsmath}
\usepackage{amssymb}
\usepackage{rotate}
\usepackage{color}
\usepackage{bm}

\DeclareFontFamily{OT1}{pzc}{}
\DeclareFontShape{OT1}{pzc}{m}{it}%
            {<-> s * [1.10] pzcmi7t}{}
\DeclareMathAlphabet{\mathscr}{OT1}{pzc}%
                                {m}{it}

\newcommand\lsim{\mathrel{\rlap{\lower4pt\hbox{\hskip1pt$\sim$}}
        \raise1pt\hbox{$<$}}}
\newcommand\gsim{\mathrel{\rlap{\lower4pt\hbox{\hskip1pt$\sim$}}
        \raise1pt\hbox{$>$}}}

\newcommand{\be}{\begin{equation}}
\newcommand{\ee}{\end{equation}}
\newcommand{\bea}{\begin{eqnarray}}
\newcommand{\eea}{\end{eqnarray}}
\def\ba#1\ea{\begin{align}#1\end{align}}

\begin{document}
\title{Note on Adiabatic Modes and Ward Identities In A Closed Universe}
\author{Xiao Xiao}
\affiliation{Department of Physics, Institute for String, Cosmology
  and Astroparticle Physics,\\ Columbia University, New York, NY 10027}

\begin{abstract}As statements regarding the soft limit of cosmological correlation functions, consistency relations are known to exist in any flat FRW universe. In this letter we explore the possibility of finding such relations in a spatially closed universe, where the soft limit $\textbf{q}\rightarrow 0$ does not exist in any rigorous sense. Despite the absence of spatial infinity of the spatial slices, we find the adiabatic modes and their associated consistency relations in a toy universe with background topology $R\times S^2$. Flat FRW universe adiabatic modes are recovered via taking the large radius limit $R\gg \mathcal{H}^{-1}$, for which we are living in a small local patch of Hubble size on the sphere. It is shown that both dilation and translation adiabatic modes in the local patch are recovered by a global dilation on the sphere, acting at different places. 
\end{abstract}
\maketitle


Consistency relations---relations between cosmological correlation functions with long wavelength operator insertions and the ones without, have been derived in the context of inflation as well as large scale structure. These relations are derived from few assumptions, while they yield highly non-trivial constraints on observational data. Being analogs for soft-pion theorems in strong interaction dynamics that were explored back in 1950's, the first of such consistency relations in cosmology appeared in the seminal work by Maldacena \cite{Malda}, it was argued and checked explicitly that for a single field inflation model, the bi-spectrum of curvature perturbation is completely determined by the spectrum, in the limit that one of the $\zeta$ modes has a vanishingly small wavenumber. The original argument by Maldacena was formalized by several authors, in \cite{bg} more general consistency relations are derived via ``background field argument''.  In \cite{HHL}\cite{HHL2} it was pointed out that the dynamics governing curvature perturbations in any flat $d+1$ dimensional FRW universe has a non-linearly realized $SO(d,2)$ conformal symmetry, with $\zeta$ interpreted as the Goldstone boson of such a conformal symmetry, consistency relations involving soft $\zeta$ mode and soft gravitational mode are derived as Ward identities. In \cite{BK}\cite{Pim}, master relations away from the soft limit are derived as consequence of gauge invariance of the quantum effective action as well as the wavefunction of universe. 

It is shown that consistency relations are relevant in the context of structure formation, in which the dark matter is approximated by a single component fluid coupled to gravity. In \cite{Keha}\cite{PP} the authors showed that the collection of equations: continuity equation, Euler's equation and Poisson's equation, that describe the Newtonian physics of structure formation, have a ``Galilean'' symmetry which implie a consistency relation constraining the correlation functions of overdensities. In \cite{Paolo}\cite{HHX}, it is clarified that the ``Galilean symmetry'' is a special case of non-linearly realized conformal symmetries, which has direct map with the ones found for curvature perturbations, further consistency relations are derived as the relativistic corrections to the ``Galilean'' relation. In \cite{HHX2}, the consistency relation derived in \cite{Keha} is shown to be equivalent to a surprisingly simple consistency relation in Lagrangian space.

All of the adiabatic modes and associated consistency relations have their origin from non--linearly realized symmetries as coordinate transformations that are not fixed by gauge--fixing conditions. These symmetries are ``large diffeomorphism'' that do not vanish at spatial infinity, while any diffeomorphism with a finite spatial momentum is fixed by gauge conditions. One may wonder if the existence of such transformations and non--trivial Ward identities can be associated to asymptotic symmetries at spatial infinity. For example in asymptotically flat spacetime we have BMS group \cite{Bondi} living at the null infinity and spi group \cite{AH} at spatial infinity, these asymptotic symmetries are large diffeomorphisms that shift the boundary data. Could the symmetries in the FRW universe find their origin at the boundary of the spatial slices and turn out to be the analog of spi group in an expanding universe?  

In this note we point out that the existence of spatial infinity is not crucial for the existence of the symmetries in cosmology which are introduced above. We find the analog of Weinberg's adiabatic mode \cite{Wei} in a universe with topology $R\times S^2$ and its associated consistency relation. The consistency relation relates angular space correlation functions with an insertion of mode carrying angular momentum $l=1$ to correlations functions without the $l=1$ mode. The correction to local consistency relations are shown to be of order $\frac{1}{\mathcal{H}^2R^2}\sim\Omega-1$.    

We work in a toy $2+1$ dimensional universe with topology $R\times S^2$. In Newtonian gauge the metric takes the form:
\begin{equation}
  ds^2=a\left( \eta \right)^2\left( -\left( 1+2\Phi \right)d\eta^2+\left( 1-2\Psi \right)R^2\left( d\chi^2+\sin^2\chi d\varphi^2 \right) \right)
  \label{}
\end{equation}
The gauge is fixed locally, however there are residual gauge transformations generated by conformal Killing vectors, here we focus on one of them which corresponds to dilation:
\begin{equation}
  \xi^\chi=\lambda\sin\chi~,~\xi^\varphi=0
  \label{dilation}
\end{equation}
This particular transformation dilates the space near the north pole $\chi=0$ and squeezes the space near the south pole. Under the transformation the metric is rescaled inhomogeneously over the sphere:
\begin{equation}
  ds^2\rightarrow \left( 1-2\cos\chi \right)ds^2
  \label{}
\end{equation}
Now we consider the analog of Weinberg's adiabatic mode \cite{Wei}, which is generated by a dilation that is constant in time and a time translation which has a non--trivial time dependence. The combined dilation and time translation induce the shifts in the gravitational potential:
\begin{equation}
  \begin{split}
 & \Phi\rightarrow \Phi-\dot{\xi}^0-\mathcal{H}\xi^0\\
 & \Psi\rightarrow \Psi+\mathcal{H}\xi^0+\partial_\chi \xi^\chi
\end{split}
  \label{shift}
\end{equation}
In the case of flat universe, the adiabatic modes are physical modes that approach the gauge modes in the limit that spatial momentum $\textbf{q}$ goes to zero.The gauge modes need to have certain time--dependence to be extrapolated away from zero spatial momentum, such adiabatic conditions \cite{Paolo}\cite{HHX}\cite{Wei} are from demanding the traceless part of the Einstein's equations
\begin{equation}
  \left( \nabla_i\nabla_j-\frac{1}{3}\gamma_{ij}\nabla^2 \right)\left( \Phi-\Psi \right)=0
  \label{}
\end{equation}
are satisfied with non--zero momentum, which means
\begin{equation}
  \Phi=\Psi.
  \label{condition}
\end{equation}
In a spherical universe we no longer have the notion of soft limit $\textbf{q}\rightarrow 0$ in a rigorous sense, since as the wave is streched it feels the spatial curvature and $\textbf{q}$ fails to be a good quantity to characterize the mode, however we expect that any adiabatic mode will reduce to adiabatic mode in a flat universe in a patch around the north pole with radius much smaller than the radius of the sphere. Thus demanding (\ref{condition}) in the closed universe, with (\ref{shift}) we arrive at the adiabatic condition:
\begin{equation}
  \dot{\xi}^0+2\mathcal{H}\xi^0+\lambda\cos\chi=0
  \label{}
\end{equation}
with solution
\begin{equation}
  \xi^0=-\frac{\lambda}{a^2\left( \eta \right)}\int^\eta d\eta'a^2\left( \eta' \right)\cos\chi
  \label{adiabatic}
\end{equation}
As we are living in a local patch near the north pole, we focus on the limit $\chi\sim 0$ and indeed the adiabatic mode in the flat FRW universe is recovered. The Eq. (adiabatic) is the generalization of flat FRW universe adiabatic mode to the case of closed universe. We see that it carries angular momentum $l=1$ on the sphere, as is true for the dilation transformation itself, thus can effectively be associated with a wavenumber $\textbf{q}\sim l \frac{1}{R}\sim\frac{1}{R}$

There is a consistency relation associated with this mode, characterizing the effects of a dilation on the modes with higher angular momentum and thus with smaller spatial scales. Since we are working on a sphere, it is convenient to decompose a given mode on spherical harmonics:
\begin{equation}
  \mathcal{O}\left( \chi,\varphi \right)=\sum_{l,m}\mathcal{O}_{l,m}Y_{lm}\left( \chi,\varphi \right)
  \label{}
\end{equation}
Under a dilation (\ref{dilation}), the linear piece in the transformation of a mode is 
\begin{equation}
  \delta\mathcal{O}_{lin.}=\lambda\sum_{l,m}\mathcal{O}_{l,m}\sin\chi\frac{\partial}{\partial\chi}Y_{lm}\left( \chi,\varphi \right)
  \label{}
\end{equation}
It is possible that a mode transforms non--linearly under the dilation and thus has a non--linear piece in its full transformation, for instance the gravitational potential $\Phi$ and $\Psi$, but it is shown in (\cite{HHL}) that to consider the consistency relations for connected correlation functions, we can just keep the linear piece.

Using the relation between spherical harmonics
\begin{equation}
  \begin{split}
  &\sin\chi\partial_\chi Y_{lm}=-\frac{1}{2l+1}[ (l+m)(l-m+1)\sqrt{\frac{(2l+1)(l-m)}{(2l-1)(l+m)}}Y_{l-1,m}\\
    &-l(l-m+1)\sqrt{\frac{(2l+1)(l+m+1)}{(2l+3)(l-m+1)}}Y_{l+1,m} ]
\end{split}
  \label{}
\end{equation}
and take the limit $l\gg 1$ while $m$ doesn't scale as $l$, we have
\begin{equation}
  \sin\chi\partial_\chi Y_{lm}\approx-\frac{l}{2}\left[ Y_{l-1,m}-Y_{l+1,m} \right]\sim l\frac{\partial}{\partial l}Y_{lm}~,~l\gg 1
  \label{}
\end{equation}
Thus on the background of a long mode corresponds to dilation, a two--point function with large angular momenta has a linear change
\begin{equation}
  \delta\langle\mathcal{O}_{l,m}\mathcal{O}_{l',m'}\rangle=-\lambda\left[ \frac{\partial}{\partial l}\left(l \langle\mathcal{O}_{l,m}\mathcal{O}_{l',m'}\rangle \right)+\frac{\partial}{\partial l'}\left( l'\langle\mathcal{O}_{l,m}\mathcal{O}_{l',m'}\rangle \right) \right]
  \label{change}
\end{equation}
The parameter $\lambda$ is identified with the angular--momentum space amplitude of the long mode via the following relation:
\begin{equation}
  \Psi_{l=1,m=0}=2\sqrt{\frac{\pi}{3}}\lambda
  \label{amp}
\end{equation}
which is from
\begin{equation}
  \Psi_{10}\left( \chi,\varphi \right)=\lambda\cos\chi=2\lambda\sqrt{\frac{\pi}{3}}Y_{10}(\chi,\varphi)
  \label{}
\end{equation}
Combining (\ref{change}) and (\ref{amp}) we have
  
\begin{equation}
  \begin{split}
  &\langle\Psi_{10}\mathcal{O}_{lm}\mathcal{O}_{l'm'}\rangle=\frac{1}{2}\sqrt{\frac{3}{\pi}}\langle\Psi_{10}\Psi_{10}\rangle\times\\
  &\left[ \frac{\partial}{\partial l}\left( l\langle\mathcal{O}_{lm}\mathcal{O}_{l'm'}\rangle\right)+\frac{\partial}{\partial l'}\left(l'\langle\mathcal{O}_{lm}\mathcal{O}_{l'm'}\rangle \right) \right]
\end{split}
  \label{}
\end{equation}
which is a relation relating a three--point function in the ``squezzed limit'' to two--point functions. Here the notion of ``squeezed limit'' is generalized from the flat space, it refers to the angular momentum configuration that contains one mode with angular momentum of order $\sim 1$ and other modes with large angular momenta $l\gg 1$. This relation is derived in a universe with spatial geometry $S^2$, to extend to cosmology on $S^3$ one can directly extend everything with $SO(4)$ spherical harmonics.

This consistency relation as it stands, is a super--horizon relation, as we assume the curvature radius of the universe is bigger than Hubble scale at least by a factor of 10 to fit the observational constraints. For observers in a small patch, the correction to local physics due to the curvature is more relevant. To get such corrections on the flat universe consistency relation, we take (\ref{dilation}) and Taylor expand around $\chi=0$:
\begin{equation}
  \chi\rightarrow \chi+\lambda\left( \chi-\frac{1}{3!R^2}\chi^3+\dots \right)
  \label{}
\end{equation}
The generated adiabatic mode from this transformation thus has a quadratic spatial dependence:
\begin{equation}
  \Psi=\lambda\left( 1-\frac{1}{2}\left( \frac{\chi}{R} \right)^2+\dots \right)
  \label{}
\end{equation}
If we describe the universe around $\chi=0$ as a flat FRW universe, then $\chi$ is identified with the radial direction from the origin. Now lt's look at the dilation consistency relation therein, with the presence of a long mode, the correlation function of short modes has a linear transformation:
\begin{equation}
  \begin{split}
    & \delta\langle\mathcal{O}(x_1)\mathcal{O}(x_2)\dots\rangle=\int\prod_b d^3\textbf{k}_b i\sum_a\mathcal{O}_{\textbf{k}_a}\cdot\delta\textbf{x}_a e^{\sum_b i\textbf{k}_b\cdot\textbf{x}_b}\\
    &=\int\prod_b d^3\textbf{k}_b i\lambda\sum_a\mathcal{O}_{\textbf{k}_a}\cdot\left(\textbf{x}_a-\frac{1}{3!R^2}\textbf{x}^2\textbf{x}_a\right) e^{\sum_b i\textbf{k}_b\cdot\textbf{x}_b}\\
  \end{split}
  \label{}
\end{equation}
Thus in momentum space a correction of order $\frac{1}{\textbf{k}^2R^2}$ is induced by the curvature, for the short modes within the Hubble scale, the correction can be up to $\sim\frac{1}{(\mathcal{H}R)^2}$:
\begin{equation}
  \begin{split}
  &\langle\Psi_{\textbf{q}\rightarrow 0}\mathcal{O}_{\textbf{k}_1}\dots\mathcal{O}_{\textbf{k}_n}\rangle=\\
  &-\langle\Psi_q\Psi_q\rangle\sum_{a=1}^n\left(\frac{\partial}{\partial\textbf{k}_a}-\frac{1}{3!R^2}\frac{\partial^2}{\partial\textbf{k}^2}\frac{\partial}{\partial\textbf{k}_a} \right)\left( \textbf{k}_a\langle\mathcal{O}_{\textbf{k}_1}\dots\mathcal{O}_{\textbf{k}_n}\rangle \right)
\end{split}
  \label{}
\end{equation}

Next we address an ambiguity arising from the identification of flat universe translation mode in the closed universe. Imagine we are living in a patch at the north pole, then a translation on our patch can be generated by a rotation around an axis which intersects with the equator. Also it can be done via a dilation which centers around a point on the equator. One can ask which of them reproduces the translation adiabatic mode in our patch. This is related to the fact that in flat space, a translation is both longitudinal and transverse, in the sense that
\begin{equation}
  \xi^i=n^i\left( \eta \right)
  \label{}
\end{equation}
is divergenceless as well as being the gradient of a scalar itself.

In a closed universe, this degeneracy is lifted: a rotation is transverse, dilation is longitudinal, and we will see that the translation mode in our patch arises from a dilation centered at the equator, not a rotation.

To see this let's look at a rotation on the sphere, to keep Newtonian gauge, the longitudinal part of the transformation in the $0i$ component of the metric
\begin{equation}
  \Delta g_{0i}=\nabla_i\xi^0-\dot{\xi}^i
  \label{0i}
\end{equation}
should vanish. For rotation, $\xi^i$ itself is transverse, satisfying 
\begin{equation}
  \nabla_i\xi^i=0
  \label{}
\end{equation}.
Thus the time translation should satisfy
\begin{equation}
  \nabla^2\xi^0=0
  \label{}
\end{equation}
In flat space this means $\xi^0$ can be a harmonic function, but on a sphere this means $\xi^0$ must carry zero angular momentum and be a constant over the whole sphere and therefore a constant on our local patch. This is not what we have in flat space, where $\xi^0$ has a linear dependence on $\textbf{X}$:
\begin{equation}
  \xi^i=n^i~,~\xi^0=\dot{n}^ix^i
  \label{tran}
\end{equation}
Then look at the dilation which is longitudinal. Now for a time--dependent dilation to respect Newtonian gauge we have demand (\ref{0i}) vanishes. Thus for a dilation centered at the north pole, we have
\begin{equation}
  \xi^\chi=\lambda\left( \eta \right)\sin\chi~,~\xi^0=-\dot{\lambda}\cos\chi
  \label{}
\end{equation}
Now let's look at a local patch on the equator, the observers living there will experience a translation. To see this we redefine the coordinate:
\begin{equation}
  \chi\equiv \frac{\pi}{2}+r
  \label{}
\end{equation}
$r$ here represents the distance of a point to the equator. Writing down the coordinate transformation with the new coordinate we see
\begin{equation}
  \begin{split}
    &\xi^r=\lambda\sin\left( \frac{\pi}{2}+r \right)\sim\lambda\\
    &\xi^0=-\dot{\lambda}\cos\left( \frac{\pi}{2}+r \right)\sim\dot{\lambda}r
  \end{split}
  \label{}
\end{equation}
which matches with (\ref{tran}) when we identify $l$ with $|\textbf{n}|$.

Thus we get a nice picture: both translation and dilation modes in a nearly--flat local patch arise from dilation on the sphere. Observers in a patch see dilation mode when the center of the dilation is at the patch, the observers see translation mode if the center of the dilation is faraway from the patch. For observers at the north pole, the dilation that produces the translation mode is centered at the equator.

In this note we see that non--linearly realized symmetries that generate adiabatic modes and their corresponding consistency relations can exist without a spatial boundary. In a closed universe such transformations generate adiabatic modes that match onto flat universe adiabatic modes for observers confined in a small Hubble size patch. All the discussion here can be generalized easily to higher dimensions, with $SO(4)$ spherical harmonics $Y_{lm_1m_2}$. Also it would be interesting to explore the symmetries corresponding to flat space special conformal transformations.

\emph{Acknowledgments:} 
This work is supported in part by United States Department of Energy under DOE grant DE-FG02-92-ER40699.


\vspace{-0.64cm}


\begin{thebibliography}{10}
  \bibitem{Malda}
    J.~M.~Maldacena,
    ``Non-Gaussian features of primordial fluctuations in single field inflationary models,''
    JHEP {\bf 0305}, 013 (2003)
    [astro-ph/0210603].
  \bibitem{bg}
    P.~Creminelli, J.~Norena and M.~Simonovic,
    ``Conformal consistency relations for single-field inflation,''
    JCAP {\bf 1207}, 052 (2012)
    [arXiv:1203.4595 [hep-th]].
  \bibitem{Keha} 
    A.~Kehagias and A.~Riotto,
    ``Symmetries and Consistency Relations in the Large Scale Structure of the Universe,''
    Nucl.\ Phys.\ B {\bf 873}, 514 (2013)
    [arXiv:1302.0130 [astro-ph.CO]].
  \bibitem{HHL}
    K.~Hinterbichler, L.~Hui and J.~Khoury,
      ``Conformal Symmetries of Adiabatic Modes in Cosmology,''
      JCAP {\bf 1208}, 017 (2012)
        [arXiv:1203.6351 [hep-th]].
  \bibitem{HHL2}
    K.~Hinterbichler, L.~Hui and J.~Khoury,
      ``An Infinite Set of Ward Identities for Adiabatic Modes in Cosmology,''
      JCAP {\bf 1401}, 039 (2014)
        [arXiv:1304.5527 [hep-th]].
  \bibitem{BK}
    L.~Berezhiani and J.~Khoury,
      ``Slavnov-Taylor Identities for Primordial Perturbations,''
      JCAP {\bf 1402}, 003 (2014)
        [arXiv:1309.4461 [hep-th]].
  \bibitem{Pim}
    G.~L.~Pimentel,
      ``Inflationary Consistency Conditions from a Wavefunctional Perspective,''
      JHEP {\bf 1402}, 124 (2014)
        [arXiv:1309.1793 [hep-th]].
  \bibitem{Paolo}
    P.~Creminelli, J.~Noreña, M.~Simonović and F.~Vernizzi,
      ``Single-Field Consistency Relations of Large Scale Structure,''
      JCAP {\bf 1312}, 025 (2013)
        [arXiv:1309.3557 [astro-ph.CO]].
  \bibitem{HHX}
    B.~Horn, L.~Hui and X.~Xiao,
    ``Soft-Pion Theorems for Large Scale Structure,''
    arXiv:1406.0842 [hep-th].
  \bibitem{HHX2}
    B.~Horn, L.~Hui and X.~Xiao,
    ``Lagrangian space consistency relation for large scale structure,''
    to appear
  \bibitem{PP}
    M.~Peloso and M.~Pietroni,
      ``Galilean invariance and the consistency relation for the nonlinear squeezed bispectrum of large scale structure,''
      JCAP {\bf 1305}, 031 (2013)
        [arXiv:1302.0223 [astro-ph.CO]].
  \bibitem{GHN}
    W.~D.~Goldberger, L.~Hui and A.~Nicolis,
      ``One-particle-irreducible consistency relations for cosmological perturbations,''
      Phys.\ Rev.\ D {\bf 87}, no. 10, 103520 (2013)
        [arXiv:1303.1193 [hep-th]].
  \bibitem{AH}
     A.~Ashtekar and R.~O.~Hansen,
       ``A unified treatment of null and spatial infinity in general relativity. I - Universal structure, asymptotic symmetries, and conserved quantities at spatial infinity,''
       J.\ Math.\ Phys.\  {\bf 19}, 1542 (1978).
  \bibitem{Bondi}
    H.~Bondi, M.~G.~J.~van der Burg and A.~W.~K.~Metzner,
      ``Gravitational waves in general relativity. 7. Waves from axisymmetric isolated systems,''
      Proc.\ Roy.\ Soc.\ Lond.\ A {\bf 269}, 21 (1962).
  \bibitem{Stro}
    T.~He, V.~Lysov, P.~Mitra and A.~Strominger,
      ``BMS supertranslations and Weinberg's soft graviton theorem,''
      arXiv:1401.7026 [hep-th].
  \bibitem{Wei}
    S.~Weinberg,
      ``Adiabatic modes in cosmology,''
      Phys.\ Rev.\ D {\bf 67}, 123504 (2003)
        [astro-ph/0302326].
\end{thebibliography}
\end{document}